\documentclass[aps,
twocolumn,
nofootinbib  % Do not show foot notes in reference list.
]{revtex4-2}

\usepackage{amsmath}
\usepackage{amssymb}
\usepackage{amsbsy}
\usepackage{amstext}
\usepackage{amsthm}
\usepackage{amsfonts}
\usepackage{physics}
\usepackage{graphicx}
\usepackage{mathrsfs}
\usepackage{hyperref}  % Hyper link.
\usepackage{soul}
\hypersetup{
	colorlinks = true,
	linkcolor = blue,
	citecolor = blue,
	urlcolor = blue
}
\usepackage{bm}
\usepackage{comment}
\usepackage[normalem]{ulem}  
\usepackage{tensor}

\usepackage{slashed}
\usepackage{simplewick}
\usepackage{enumitem}
\usepackage{lipsum}

\def\l{\left}
\def\r{\right}
\def\({\l(}
\def\){\r)}
\def\[{\l[}
\def\]{\r]}
\def\la{\langle}
\def\ra{\rangle}
\newcommand{\pvec}{{\bm p}}

\def\rm{\mathrm}
\def\cal{\mathcal}
\def\pd{\partial}

\def\b{\boldsymbol}

\def\bar{\overline}
\def\sl{\slashed}  % Feynman slash.

\def\j{\varphi}

\def\ve{\varepsilon}

\def\sep{~,\quad}  % Separate equations.

\newcommand{\yr}{\, \mathrm{yr}}

\newcommand{\Kel}{\, \mathrm{K}}
\newcommand{\ueV}{\, \mu\mathrm{eV}}
\newcommand{\keV}{\, \mathrm{keV}}
\newcommand{\MeV}{\, \mathrm{MeV}}
\newcommand{\GeV}{\, \mathrm{GeV}}

\begin{document}
\title{Neutron star cooling with lepton-flavor-violating axions}  % Force line breaks with \\
\author{Hong-Yi Zhang}
\email{hongyi@rice.edu}
\author{Ray Hagimoto}
\email{ray.m.hagimoto@rice.edu}
\author{Andrew J. Long}
\affiliation{Department of Physics and Astronomy, Rice University, Houston, Texas 77005, USA}
\date{\today}
\begin{abstract}
The cores of dense stars are a powerful laboratory for studying feebly coupled particles such as axions. Some of the strongest constraints on axionlike particles and their couplings to ordinary matter derive from considerations of stellar axion emission. In this work we study the radiation of axionlike particles from degenerate neutron star matter via a lepton-flavor-violating coupling that leads to muon-electron conversion when an axion is emitted. We calculate the axion emission rate per unit volume (emissivity) and by comparing with the rate of neutrino emission, we infer upper limits on the lepton-flavor-violating coupling that are at the level of $|g_{ae\mu}| \lesssim 10^{-6}$. For the hotter environment of a supernova, such as SN 1987A, the axion emission rate is enhanced and the limit is stronger, at the level of $|g_{ae\mu}| \lesssim 10^{-11}$, competitive with laboratory limits. Interestingly, our derivation of the axion emissivity reveals that axion emission via the lepton-flavor-violating  coupling is suppressed relative to the familiar lepton-flavor-preserving channels by the square of the plasma temperature to muon mass ratio, which is responsible for the relatively weaker limits.
\end{abstract}
\maketitle

%========================================
%========================================
\section{Introduction}
Axions are pseudo-Goldstone bosons associated with a spontaneously broken global symmetry that is anomalous to the standard model (SM) gauge couplings \cite{Kim:1986ax}. Initially proposed as a natural solution to explain the absence of the neutron electric dipole moment \cite{Peccei:1977hh, Weinberg:1977ma, Wilczek:1977pj}, a QCD axion is characterized by its decay constant $f_a$ \cite{Kim:1979if, Shifman:1979if, Zhitnitsky:1980tq, Dine:1981rt} and its mass is determined by $m_a \approx 5.7 \ueV (10^{12}\GeV/f_a)$ \cite{Borsanyi:2016ksw, Gorghetto:2018ocs}. Apart from the QCD axion, axionlike particles have also been extensively studied in string theory \cite{Svrcek:2006yi, Arvanitaki:2009fg, Ringwald:2012cu} and dark matter physics \cite{Dine:1982ah, Abbott:1982af, Preskill:1982cy, Ferreira:2020fam, Hui:2021tkt}. For recent reviews, refer to \cite{Marsh:2015xka, DiLuzio:2020wdo, Sikivie:2020zpn, ParticleDataGroup:2022pth}.

Due to their weak interactions with SM particles, detecting axions in terrestrial experiments is challenging. Therefore, it is motivated to search for evidence of axions in astrophysical systems where their feeble couplings are partially compensated by high temperatures and densities \cite{Raffelt:1990yz}. For instance, probing axion emission from the white dwarf luminosity function \cite{MillerBertolami:2014rka, Giannotti:2017hny, Isern_2018, Corsico:2019nmr} places a stringent limit on the axion-electron coupling at the level of $g_{aee} \lesssim 10^{-13}$.  Additionally, the axion's interaction with nucleons is probed by neutron star (NS) cooling \cite{Hamaguchi:2018oqw, Beznogov:2018fda, Buschmann:2021juv} and supernova neutrino emission \cite{Burrows:1988ah, Burrows:1990pk, Keil:1996ju, Hanhart:2000ae, Fischer:2016cyd, Carenza:2019pxu, Carenza:2020cis, Lella:2023bfb}, which imply tight upper limits at the level of $g_{aNN} \lesssim 10^{-10}$. 

As an extension of the SM, there is no strong reason for the ultraviolet theory of axions to respect lepton flavor conservation since it is an accidental symmetry of the SM broken by tiny neutrino masses. The axions whose ultraviolet theory is responsible for the breaking of the flavor symmetry are known as flavons or familons \cite{Davidson:1981zd, Wilczek:1982rv, Anselm:1985bp, Feng:1997tn, Bauer:2016rxs}, which can also explain the strong CP problem if they have a coupling to gluons \cite{Ema:2016ops, Calibbi:2016hwq}. Even if the underlying theory preserves lepton flavor, lepton-flavor-violating (LFV) effects can arise from radiative corrections \cite{Choi:2017gpf, Chala:2020wvs, Bauer:2020jbp, Bonilla:2021ufe}. It has been shown that LFV interactions can account for the production of dark matter through thermal freeze-in \cite{Panci:2022wlc}. Tests of lepton flavor conservation thus provide important information about new physics.

\begin{figure}
	\centering
	\includegraphics[width=0.6\linewidth]{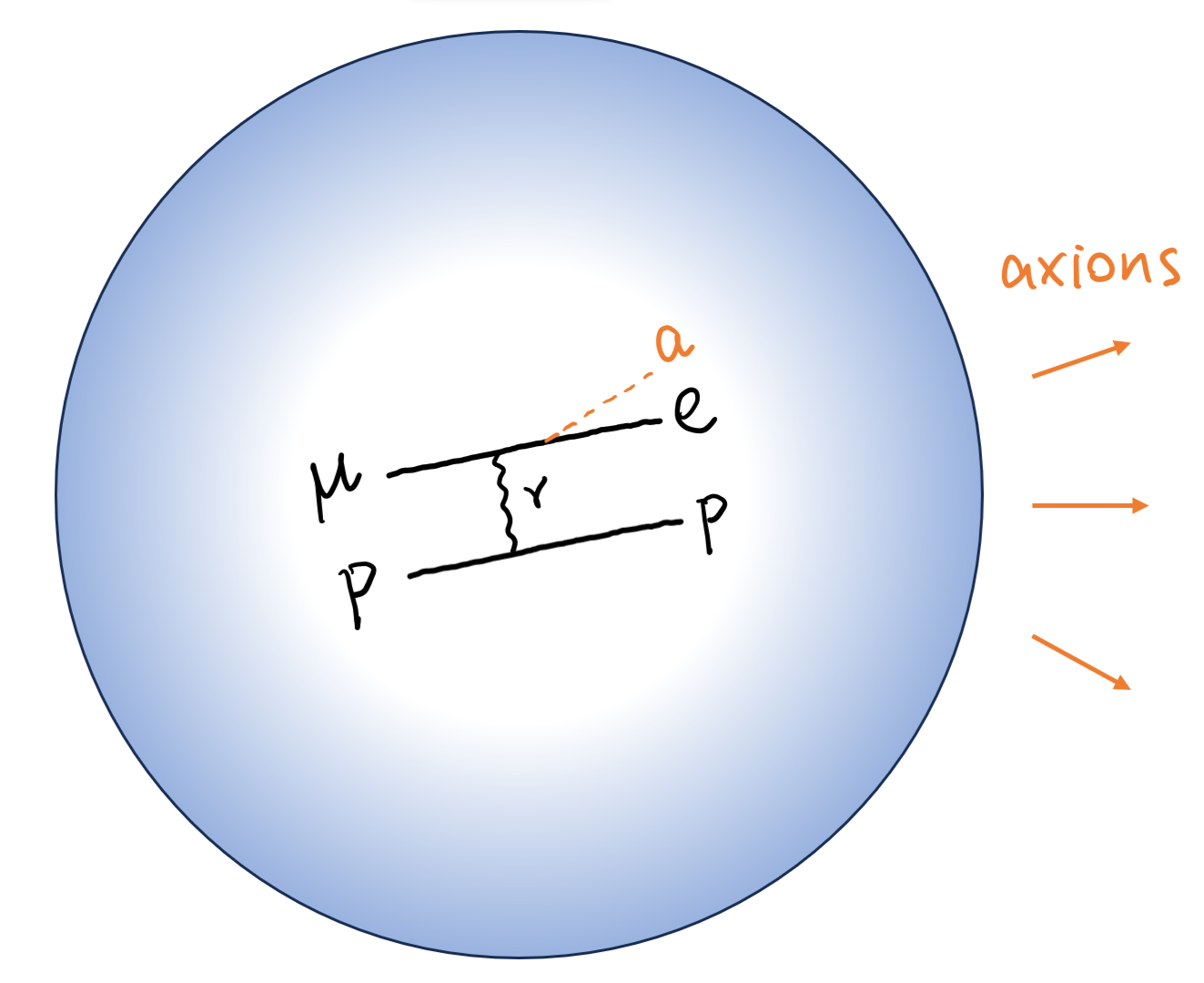}
	\caption{If axions are produced in neutron star cores, they will carry energy out of the star and make the neutron star cool down more efficiently than expected.}
	\label{fig:lfvemission}
\end{figure}

Laboratory tests of lepton-flavor violation serve as an indirect probe of the axion's LFV interactions. Notably, charged lepton flavor violation would lead to rare lepton decays \cite{Calibbi:2017uvl}. If the axion were heavier than the muon, an effective field theory approach could be used to study decays such as $\mu \rightarrow e \gamma$, $\mu \rightarrow 3e$ and $\mu-e$ conversion, being the best process to detect LFV in the $e\mu$ sector.\footnote{In the SM, LFV decays are suppressed by the neutrino mass-squared difference and $\rm{Br}(\mu\rightarrow e\gamma) \sim \rm{Br}(\mu\rightarrow 3e) \sim 10^{-54}$ \cite{Petcov:1976ff, Calibbi:2017uvl, Hernandez-Tome:2018fbq}, far below the current experimental limits $\rm{Br}(\mu\rightarrow e\gamma) < 4.2 \times 10^{-13}$ \cite{MEG:2016leq} and $\rm{Br}(\mu\rightarrow 3e) < 1.0\times 10^{-12}$ \cite{SINDRUM:1987nra}.} For lighter axions, $\mu\rightarrow ea$ could be the dominating channel and the current limit on $\rm{Br}(\mu\rightarrow ea)$ is of order $10^{-6}$ \cite{Jodidio:1986mz} or $10^{-5}$ \cite{TWIST:2014ymv} depending on the axion mass and chirality of the interaction. The limit will be improved in the future experiments MEG II \cite{MEGII:2018kmf, Jho:2022snj} and Mu3e \cite{Perrevoort:2018ttp} by up to two orders of magnitude \cite{Calibbi:2020jvd}.

In this work, we aim to establish an astrophysical limit on the axion's LFV interactions based on NS cooling arguments, as a complement to current lab limits. The basic idea is illustrated in figure \ref{fig:lfvemission}; if axions are produced in NS cores, they must not carry energy out of the star more efficiently than standard neutrino-mediated cooling channels \cite{Raffelt:1990yz}. In a NS core, unlike nondegenerate stars or even white dwarf stars, the particle densities are so high that the electron Fermi energy exceeds the muon mass, and an appreciable population of muons is present \cite{Haensel:2007yy}. As such, NSs provide a unique opportunity to probe the axion's LFV coupling with muons and electrons.

%========================================
%========================================
\section{Axions with LFV couplings}
We consider a LFV coupling among the electron, muon, and axion, which is expressed as
\begin{align}
	\label{lfv}
    \mathscr{L}_\rm{LFV} = \frac{g_{ae\mu}}{m_e + m_\mu} \bar\Psi_e \gamma^\rho \gamma_5 \Psi_\mu \, \pd_\rho a \ + \ \mathrm{h.c.} ~,
\end{align}
where $\Psi_e(x)$ is the electron field, $\Psi_\mu(x)$ is the muon field, $a(x)$ is the axion field, $m_e \approx 0.511 \MeV$ is the electron mass, $m_\mu \approx 106 \MeV$ is the muon mass, and $g_{ae\mu}$ is the axion's LFV coupling. The coupling may also be written in terms of the axion decay constant $f_a$ as $g_{ae\mu} = C_{ae\mu}(m_e + m_\mu)/(2f_a)$. This interaction can naturally arise, e.g., in the models of the LFV QCD axion \cite{Zhitnitsky:1980tq, Dine:1981rt}, the LFV axiflavon \cite{Calibbi:2016hwq, Ema:2016ops, Linster:2018avp}, the leptonic familon \cite{Froggatt:1978nt, Leurer:1992wg, Leurer:1993gy} and the majoron \cite{Chikashige:1980ui, Schechter:1981cv} (also see \cite{Calibbi:2020jvd} for a summary of constraints). Past studies of charged lepton flavor violation, from both terrestrial experiments and cosmological / astrophysical observations, furnish constraints on the axion LFV coupling $g_{ae\mu}$, which we summarize here.

The LFV interaction opens an exotic decay channel for the muon $\mu\rightarrow ea$, as long as the axion mass is not too large $m_a < m_\mu - m_e$.  The branching ratio is predicted to be \cite{Bjorkeroth:2018dzu}
\begin{align}
	\mathrm{Br}(\mu \rightarrow ea) \approx \frac{\Gamma(\mu\rightarrow ea)}{\Gamma(\mu\rightarrow e \nu \bar\nu)} = 7.0 \times 10^{15} g_{ae\mu}^2 ~.
\end{align}
Initial searches for the two-body muon decay were performed by Derenzo using a magnetic spectrometer, resulting in an upper limit on the branching ratio of $2 \times 10^{-4}$ for the mass range $98.1$--$103.5 \MeV$ \cite{Derenzo:1969za}. Jodidio et al. constrained the branching ratio for a massless familon to be $< 2.6\times 10^{-6}$, which was later extended to massive particles up to $\sim 10 \MeV$ \cite{Calibbi:2020jvd}. Bryman and Clifford analyzed data of muon and tauon decays obtained from NaI(Tl) and magnetic spectrometers, concluding an upper limit of $3 \times 10^{-4}$ for masses less than $104 \MeV$ \cite{Bryman:1986wn}. Bilger et al. studied muon decay in the mass range $103$--$105 \MeV$ using a high purity germanium detector and established a limit of $5.7 \times 10^{-4}$ \cite{Bilger:1998rp}, while the PIENU Collaboration improved the limit in the mass range $87.0$--$95.1 \MeV$ \cite{PIENU:2020loi}. The TWIST experiment performed a broader search for masses up to $\sim 80 \MeV$ by accommodating nonzero anisotropies, resulting in an upper limit of $2.1\times 10^{-5}$ for massless axions \cite{TWIST:2014ymv}. These constraints on $\mathrm{Br}(\mu \rightarrow ea)$ translate into upper limits on the LFV coupling $g_{ae\mu}$, and we summarize the current status in table \ref{tab:lfv_constraint}.
\begin{table*}
	\centering
	\begin{tabular}{|c|c|c|c|c|c|}
		\hline
		$|g_{ae\mu}|$ & $\frac{2f_a}{C_{ae\mu}} ~[\rm{GeV}]$ & $\rm{Br}(\mu\rightarrow ea)$ & $m_a ~ [\rm{MeV}]$ & Experiment & Reference\\
		\hline
		$<3.0 \times 10^{-6}$ & $>3.5\times 10^4$ & $<1.0$ & $\lesssim 1$ & NS cooling & This work \\
		\hline
		$\lesssim 8\times 10^{-10}$ & $\gtrsim 1\times 10^8$ & $\lesssim 4 \times 10^{-3}$ & $ \lesssim 50$ & SN 1987A, $\mu\rightarrow ea$ & \cite{Calibbi:2020jvd} \\
		\hline
		$<4.2\times 10^{-10}$ & $>2.5\times 10^8$ & $<1.3 \times 10^{-3}$ & $\lesssim 10^{-7}$ & Cosmology, $\Delta N_\rm{eff}$ & \cite{DEramo:2021usm} \\
		\hline
		$<2.9\times 10^{-10}$ & $>3.7\times 10^8$ & $<5.7 \times 10^{-4}$ & $103-105$ & Rare muon decay & \cite{Bilger:1998rp} \\
		\hline
		$\lesssim 2\times 10^{-10}$ & $\gtrsim 5\times 10^8$ & $\lesssim 3 \times 10^{-4}$ & $<104$ & Rare muon decay & \cite{Bryman:1986wn} \\
		\hline
		$<2\times 10^{-10}$ & $>6\times 10^8$ & $<2 \times 10^{-4}$ & $98.1 - 103.5$ & Rare muon decay & \cite{Derenzo:1969za} \\
		\hline
		$<1\times 10^{-10}$ & $>9\times 10^8$ & $< 1\times 10^{-4}$ & $47.8 - 95.1$ & Rare muon decay (PIENU)\footnote{The PIENU Collaboration obtained upper limits on the branching ratio from $10^{-4}$ to $10^{-5}$ for the considered mass range.} & \cite{PIENU:2020loi} \\
		\hline
		$<5.5\times 10^{-11}$ & $>1.9\times 10^9$ & $<2.1 \times 10^{-5}$ & $< 13$ & Rare muon decay (TWIST) & \cite{TWIST:2014ymv} \\
		\hline
		$\lesssim 4\times 10^{-11}$ & $\gtrsim 3\times 10^9$ & $\lesssim 9 \times 10^{-6}$ & $ \lesssim 50$ & SN 1987A, $lf\rightarrow l'fa$ & This work \\
		\hline
		$<1.9\times 10^{-11}$ & $>5.5\times 10^9$ & $<2.6 \times 10^{-6}$ & $\lesssim 10$ & Rare muon decay & \cite{Jodidio:1986mz, Calibbi:2020jvd} \\
		\hline
	\end{tabular}
	\caption{Summary of constraints on the axion's LFV coupling in the $e$-$\mu$ sector, where stronger constraints are presented at the bottom. See the main text for more detailed descriptions. For the NS cooling limit, we calculate the axion emissivity via $l+f \to l' + f + a$ and compare with the neutrino emissivity via Murca channels. For the SN 1987A limit, we compare with the upper bound on energy loss rate.}
	\label{tab:lfv_constraint}
\end{table*}

Apart from terrestrial experiments, cosmological and astrophysical observations also constrain the axion's LFV interaction. If this interaction were too strong, relativistic axions would be produced thermally in the early universe; however, the presence of a dark radiation in the universe is incompatible with observations of the cosmic microwave background anisotropies. Constraints on dark radiation are typically expressed in terms of a parameter $N_\mathrm{eff}$ called the effective number of neutrino species.  A recent study of flavor-violating axions in the early universe finds that current observational limits on $N_\mathrm{eff}$ require the LFV coupling to obey $|2f_a/C_{ae\mu}| > 2.5 \times 10^{8} \rm{GeV}$ \cite{DEramo:2021usm}. Astrophysical probes of the axion's LFV interaction have not been extensively explored.  Calibbi et al. considered the bound on $\mathrm{Br}(\mu \to ea)$ from SN 1987A associated with the cooling of the proto-NS \cite{Calibbi:2020jvd}.  Assuming that the dominant energy loss channel is free muon decay $\mu \to ea$, they derive an upper limit on the branching ratio at the level of $4\times 10^{-3}$. We find that a stronger constraint is obtained from the 2-to-3 scattering channels, such as $\mu p \to e p a$, and we discuss this result further below.

To provide a comprehensive overview, we also introduce the constraints on LFV couplings involving $\tau$ leptons. Currently, laboratory limits on the branching ratios of rare tauon decays are $\rm{Br}(\tau\rightarrow ea)<2.7\times 10^{-3}$ and $\rm{Br}(\tau\rightarrow \mu a) < 4.5\times 10^{-3}$ \cite{ARGUS:1995bjh, Calibbi:2020jvd}. Constraints from $N_\rm{eff}$ are more stringent, $\rm{Br}(\tau\rightarrow ea)\lesssim 3\times 10^{-4}$ and $\rm{Br}(\tau\rightarrow \mu a) \lesssim 5\times 10^{-4}$ \cite{DEramo:2021usm}. Each of these limits is expected to improve significantly, by up to three orders of magnitude, in the future Belle II \cite{Belle-II:2018jsg, Calibbi:2020jvd} and CMB-S4 experiment \cite{CMB-S4:2016ple, Abazajian:2019eic, DEramo:2021usm}. However, it remains challenging to impose constraints on $\tau$ leptons from astrophysical systems due to their considerable mass of $1.8 \GeV$, which far exceeds stellar core temperatures.

%========================================
%========================================
\section{Axion emission via LFV couplings}
The emission of axions from NS matter via the LFV interaction can proceed through various channels. One might expect the dominant channel to be the decay of free muons $\mu \to ea$; however, since the electrons in NS matter are degenerate, this channel is Pauli blocked, and its rate is suppressed in comparison with scattering channels. Since NS matter consists of degenerate electrons, muons, protons, and neutrons, various scattering channels are available. We denote these collectively as\footnote{We neglect the Compton process for axions, since the number density of photons is low compared to other particles.}
\begin{align}
	\label{lfv1}
	l + f \rightarrow l' + f + a ~,
\end{align}
where a lepton $l=e,\mu$ is converted to another $l'=\mu,e$ with the spectator particle $f=p, e, \mu$. We consider channels in which the NS's muon is present in the initial state, and channels in which muons are created thanks to the large electron Fermi momentum. The scattering is mediated by the electromagnetic interaction (photon exchange), and channels involving neutrons are neglected. Assuming that all particles are degenerate, scattering predominantly happens for particles at the Fermi surface. These processes are kinematically allowed if $|p_{F,l} - p_{F,f}| < p_{F,l'} + p_{F,f}$ and $|p_{F,l'} - p_{F,f}| < p_{F,l} + p_{F,f}$, implying the existence of a threshold momentum of the spectator particle
\begin{align}
	\label{threshold}
	p_{F,f} > (p_{F,e} - p_{F,\mu}) / 2 ~.
\end{align}
Here we have introduced the Fermi momentum $p_{F,i}$ of the particle species $i$.

The quantities of interest are the axion emissivities $\ve_a^{(lf)}$, which corresponds to the energy released in axions per unit volume per unit time through the channel $lf \to l^\prime f a$. We assign $(E_1,\b p_1)$ and $(E_1',\b p_1')$ for the initial and final four-momenta of the converting leptons $l$ and $l'$, $(E_2,\b p_2)$ and $(E_2',\b p_2')$ for the spectator $f$, and $(E_3',\b p_3')$ for the axion. Then the axion emissivity is calculated as
\begin{equation}
\begin{split}
	\label{emissivity_integral}
	\ve_a^{(lf)} & = \frac{(2\pi)^4}{S} \! \int 
    \prod_{i=1}^2 \widetilde{dp_i} \ 
    \prod_{j=1}^3 \widetilde{dp_j'} \ 
    \sum\limits_\rm{spin} \, \bigl| \mathcal{M}^{(lf)} \bigr|^2 
    \\ & \quad 
    \times \delta^{(4)}(p_1+p_2-p_1'-p_2'-p_3') 
    \\ & \quad 
	\times E_3' \, f_1 \, f_2 \, (1-f_1') \, (1-f_2') ~,
\end{split}
\end{equation}
where $S$ is the symmetry factor accounting for identical initial and final state particles, $\mathcal{M}^{(lf)}$ is the Lorentz invariant matrix element, $f_i$ and $f_i'$ are the Fermi-Dirac distribution functions, the factor $(1-f_i')$ takes into account the Pauli blocking due to particle degeneracy, and $\widetilde{dp}\equiv d^3 p/[(2\pi)^3 2E]$ is the Lorentz-invariant differential phase space element. We do not include a factor of $(1+f_3')$, since $f_3' \ll 1$ and there is no Bose enhancement for axion production since NSs are essentially transparent to axions for the currently allowed parameter space.

Calculating the emissivity \eqref{emissivity_integral} requires evaluating the 15 momentum integrals along with the 4 constraints from energy and momentum conservation.  We evaluate all but 2 of these integrals analytically using the Fermi surface approximation, and we calculate the last 2 integrals using numerical techniques. The Fermi surface approximation assumes that the integrals are dominated by momenta near the Fermi surface $|{\bm p}| \approx p_F$; smaller and larger momenta do not contribute because of Pauli blocking or Boltzmann suppression. We find the axion emissivity of the $lf \to l'fa$ channel to be
\begin{align}
	\label{emissivity}
	\ve_a^{(lf)} = \frac{328\pi^2 \alpha^2 g_{ae\mu}^2}{945 m_\mu^4} \frac{\beta_{F,l} E_{F,e}^3}{\beta_{F,f}^2 p_{F,f}^2} F^{(lf)} T^8 ~,
\end{align}
where $\alpha \approx 1/137$ is the electromagnetic fine-structure constant, $E_{F,i}$ is the Fermi energy, $\beta_{F,i}\equiv p_{F,i}/E_{F,i}$ is the Fermi velocity, $T$ is the plasma temperature, and $F^{(lf)}$ is a factor depending on both the specific process and the Fermi velocity of the scattering particles. To derive \eqref{emissivity}, we have assumed that the axion mass is small compared to the NS temperature $m_a\ll T$, muons and electrons are in the beta equilibrium (i.e., $E_{F,e} \approx E_{F,\mu}$), electrons are ultra relativistic but muons are not (i.e., $p_{F,\mu}\lesssim m_\mu$), and $T\ll m_\mu^2/E_{F,e}$. Our derivation of \eqref{emissivity} can be found in appendix \ref{app:calculation_detail}.  In addition, we evaluate the emissivity fully numerically using Monte Carlo integration methods to estimate the integrals in \eqref{emissivity_integral} without employing the Fermi surface approximation.  In the regime of interest, the two methods agree very well. The impact of an axion mass $m_a\gtrsim T$ is discussed in appendix \ref{app:numerical-integration}.

The temperature dependence of the axion emissivity \eqref{emissivity} is especially interesting and important for understanding the limits from NS cooling.  For comparison, note that axion bremstrahlung via lepton-flavor-preserving (LFP) interactions (such as $ep \to epa$ or $\mu p \to \mu p a$) goes as $\ve_a \propto T^6$.  In other words, the LFV interaction leads to an emissivity that's suppressed by an additional factor of $T^2 E_{F,e}^2 / (m_\mu^2 - m_e^2)^2 \sim T^2/m_\mu^2$, which is of order $(100 \keV/100 \MeV)^2 \sim 10^{-6}$ for $T\sim 10^{9}\,\rm{K}$. A detailed discussion appears in appendix \ref{app:calculation_detail}, but the essential idea can be understood as follows. The phase-space integrals over momenta can be converted to energy integrals, and each integral for degenerate leptons and protons is restricted to the Fermi surface of thickness $\sim T$, giving a factor of $T^4$. The phase-space integral of axions (i.e., $d^3p_3'/E_3'$) gives a factor of $T^2$. The axions are emitted thermally and have an energy $\sim T$. The energy conservation delta function gives $T^{-1}$. The squared matrix element has a temperature dependence $T^2$. Putting all these together, we see that the emissivity is proportional to $T^8$. In comparison, the squared matrix element for the LFP interactions has no temperature dependence since one power of $T$ from the coupling vertex is canceled by $T^{-1}$ from the lepton propagator.

\begin{figure}
	\centering
	\includegraphics[width=0.9\linewidth]{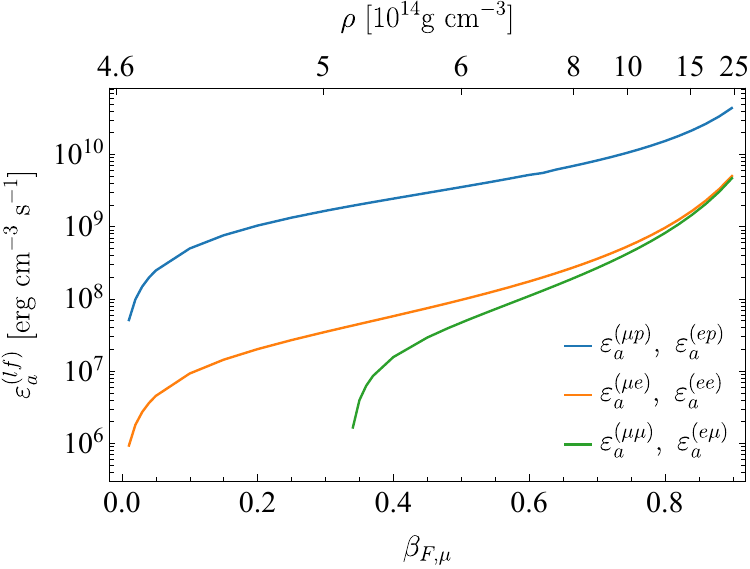}
	\caption{Axion emissivities $\ve_a^{(lf)}$ for the LFV process $l+f\rightarrow l' + f + a$, given by equation \eqref{emissivity}, as a function of the muon Fermi velocity $\beta_{F,\mu}$. The top axis, in a nonlinear scale, represents the corresponding mass density of a NS assuming the $npe\mu$ matter. Here we take $g_{ae\mu}=10^{-11}$ and $T=10^9\Kel$, and more generally $\ve_a^{(lf)} \propto g_{ae\mu}^2 T^8$.}
	\label{fig:lfvemissivity}
\end{figure}

We numerically evaluate the axion emissivities \eqref{emissivity} and present these results in figure \ref{fig:lfvemissivity} for the six channels $lf \to l'fa$, where the effective mass of protons is taken to be $0.8 \, m_p$ (see \cite{Li:2018lpy} and references therein).\footnote{Using electric charge neutrality and the beta equilibrium condition $E_{F,e}\approx E_{F,\mu}$, the emissivity is fully determined given the effective proton mass and $\beta_{F,\mu}$.} Using the strong degeneracy of particles and the beta equilibrium condition $E_{F,e}\approx E_{F,\mu}$, one can show that the emissivities are equal for the channels $ef \to \mu fa$ and $\mu f \to efa$. Thus the plot only shows three curves corresponding to in-states consisting of a muon and a spectator particle $f=p,e,\mu$. The channels with a spectator proton ($f=p$) have the largest emissivity across the range of muon Fermi momenta shown here; this is a consequence of the enhanced matrix element and the larger available phase space for these scatterings. For the channels with a spectator muon ($f=\mu$), the emissivity drops to zero below $\beta_{F,\mu} \approx 0.34$; this corresponds to a violation of the kinematic threshold in \eqref{threshold}.  For all channels, the emissivity decreases with decreasing muon Fermi velocity due to the reduced kinematically allowed phase space. On the other hand, for larger muon Fermi velocity, the channels with spectator electrons and muons coincide, since both particles can be regarded as massless. For the top axis in figure \ref{fig:lfvemission}, we show the corresponding mass density of a NS assuming the $npe\mu$ model; see appendix \ref{app:npeu_matter} for more details.

The total axion emissivity is obtained by summing over the six channels.  For this estimate we set $\beta_{F,\mu}=0.84$. 
We find the axion emissivity via LFV interactions to be
\begin{align}
	\label{emissivity_lfv}
	\ve_a^\rm{LFV} \simeq 4.8\times 10^{32} g_{ae\mu}^2 T_9^8 ~\rm{erg~cm^{-3}~s^{-1}} ~,
\end{align}
where $T_9\equiv T/(10^9\Kel)$ and $10^9\Kel \approx 86.2 \keV$.

%========================================
%========================================
\section{Implications for NS cooling}
In low-mass NSs, slow cooling could occur via neutrino emission by the modified Urca (Murca) processes $nn\rightarrow npe\bar\nu$, $npe\rightarrow nn\nu$ or slightly less efficient processes such as the nucleon bremsstrahlung \cite{Yakovlev:2004iq, Potekhin:2015qsa}. At the density $\rho=6\rho_0$, where $\rho_0=2.5\times 10^{14} \, \rm{g~cm^{-3}}$ is the nuclear saturation density \cite{Horowitz:2020evx}, and with the effective nucleon mass taken to be $0.8 \, m_N$ \cite{Li:2018lpy}, the emissivity of the Murca process is given by $\ve_\nu = 4.4 \times 10^{21} T_9^8 ~ \rm{erg~cm^{-3}~s^{-1}}$ \cite{Friman:1979ecl}. Comparing this rate with \eqref{emissivity_lfv}, one finds that the axion emission from LFV couplings dominates the neutrino emission unless
\begin{align}
    |g_{ae\mu}| \lesssim 3.0 \times 10^{-6} ~,
\end{align}
which is consistent with existing constraints. In heavier NSs, the LFV emission of axions tends to have a less significant impact. This is because fast neutrino emission could occur via the direct Urca processes \cite{Lattimer:1991ib}. In the presence of superfluidity, the formation of Cooper pairs can dominate over the Murca process \cite{Leinson:2009mq, Leinson:2009nu}, further diminishing the role of LFV axion emission. 
Medium effects for neutrino emission processes are discussed in \cite{Potekhin:2015qsa, Voskresensky:1986af, Grigorian:2016leu}.

Axions are predominantly produced in NSs through the nucleon bremsstrahlung process $nn\rightarrow nna$. At the same core conditions, its emissivity is given by $\ve_a^{(nn)} \simeq 2.8\times 10^{38} g_{ann}^2 T_9^6 ~\rm{erg~cm^{-3}~s^{-1}}$ \cite{Iwamoto:1984ir, Brinkmann:1988vi}. The nucleon bremsstrahlung process dominate the LFV processes if 
\begin{align}
    |g_{ae\mu}| \lesssim 7.6 \times 10^2 |g_{ann}| T_9^{-1} ~.
\end{align}
The current best constraint on the axion-neutron coupling is $|g_{ann}|\lesssim 2.8\times 10^{-10}$ \cite{Beznogov:2018fda}. Therefore, it is unlikely for the LFV couplings to play a significant role in NSs with an age $\gtrsim 1 \yr$, where the temperature has cooled to $10^9 \Kel$ \cite{Pethick:1991mk}. 

These limits on the axion's LFV coupling are relatively weak, and this is a consequence of the $\ve_a^\rm{LFV} \propto T^8$ scaling, which is suppressed compared to LFP channels by a factor of $(T/m_\mu)^2$, which is tiny in old NSs. However, in the proto-NS that forms just after a supernova, this ratio can be order one, which suggests that stronger limits can be obtained by considering the effect of axion emission on supernova rather than NSs.  Since our analysis has focused on NS environments, adapting our results to the more complex proto-NS system requires some extrapolation.  We estimate the axion emissivity from a supernova by extrapolating \eqref{emissivity_lfv} to high temperatures. By imposing the bound on the energy loss of SN 1987A, $\ve_a/\rho \lesssim 10^{19} \, \rm{erg~g^{-1}~s^{-1}}$ \cite{Raffelt:1990yz}, one finds that at a typical core condition $\rho\sim 8\times 10^{14} \, \rm{g~cm^{-3}}$,
\begin{align}
	\label{constraint_sn}
	\abs{g_{ae\mu}} \lesssim 4\times 10^{-11} \bigg( \frac{50 \MeV}{T} \bigg)^4 ~,
\end{align}
which is to be evaluated at $T\sim (30-60) ~\MeV$.
This constraint is more stringent than that obtained from considering $\mu\rightarrow ea$ in a supernova and is comparable to the current best terrestrial limit. 

One should note that at typical core conditions of a proto-NS, nucleons and muons are at the borderline between degeneracy and nondegeneracy where electromagnetic field screening effects become significant. In appendix \ref{app:numerical-integration}, we discuss the effect of electromagnetic field screening due to the presence of a degenerate plasma with charged constituents on the axion emissivity. We then account for this effect in our numerical code by introducing an effective mass for photon propagators of order the Thomas-Fermi wavenumber $k_{\rm{TF}}$.\footnote{While this methodology is not apt for strongly coupled plasmas like NSs and white dwarfs, it does furnish reasonably accurate estimates of the screening effect in axion bremsstrahlung processes within white dwarfs \cite{Raffelt:1990yz}.} 
Using Monte Carlo integration we evaluate the axion emissivity up to temperatures of $100~\MeV$ and find that extrapolating the degenerate rate tends to overestimate the emissivity by a factor of $\sim 10$, leading to a weaker supernova constraint by a factor $\sim 3$.

%========================================
%========================================
\section{Discussion}
\begin{figure}
    \centering
    \includegraphics[width=\linewidth]{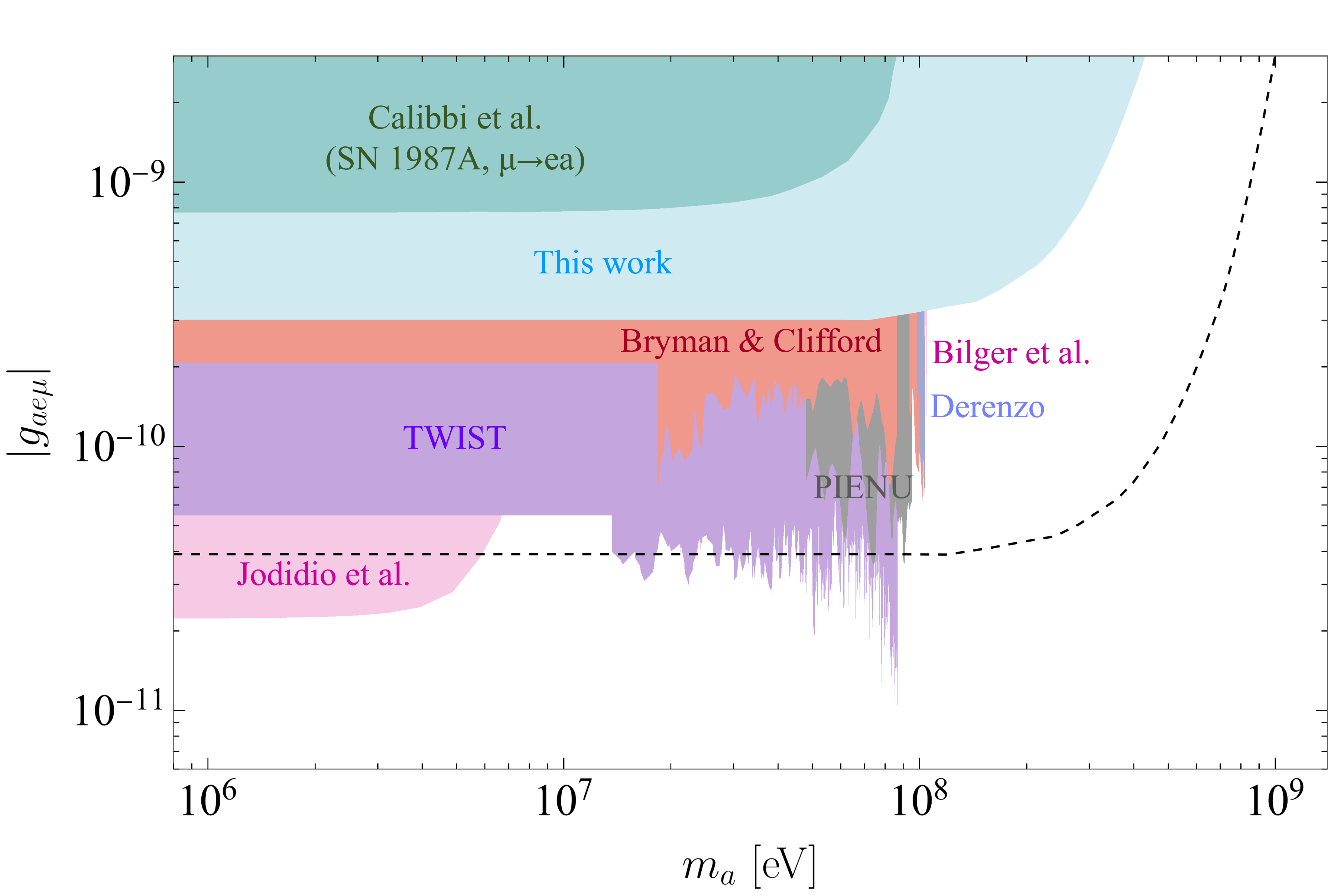}
    \caption{Summary of constraints on the axion's LFV coupling in the $e$-$\mu$ sector. The constraints labeled with ``Calibbi et al.'' and ``This work'' are astrophysical and the others are lab limits obtained by measuring rare muon decay rates. The weaker constraint we derive from NS cooling and the cosmological constraint inferred from the $\Delta N_\rm{eff}$ observation, shown in table \ref{tab:lfv_constraint}, do not appear on this part of parameter space. For the region labeled with ``This work'', we assume a supernova core temperature $T=30 \MeV$ and a higher temperature $T=50 \MeV$ would expand the exclusion region into that enclosed by the black dashed line.}
    \label{fig:lfvconstraintsummary}
\end{figure}

In this article, we study the astrophysical signatures of an axionlike particle's LFV coupling with muons and electrons. We focus on axion emission from NS cores, where the electron Fermi energy is large enough to maintain a high abundance of muons. Our limits on the LFV coupling $g_{ae\mu}$ derive from comparing the axion emission rate with the energy loss rate due to neutrino emission, since excessively strong axion emission would conflict with the observations of old NSs and SN 1987A. The summary of current constraints is shown in figure \ref{fig:lfvconstraintsummary}.

Further research is needed to assess the impact of axion's LFV interactions on the entire cooling history of the star, including a careful treatment of equations of state and nuclear interactions. Stronger nuclear interactions would result in higher number densities of protons and muons at the same mass density, thereby enhancing the rate of the LFV interactions. Such an analysis is particularly motivated for axion emission from proto-NSs formed after type-II supernovae, where the transition from nondegenerate to degenerate matter and the creation of the muon population could impact axion emissivities. Our work highlights the importance of assessing both the free muon decay channel $\mu \to e a$ as well as scattering channels $lf \to l'fa$ in such studies. 

%========================================
%========================================
\begin{acknowledgments}
	We would like to thank Mustafa Amin for helpful discussions. H.Y.Z. is partly supported through DOE Grant No. DESC0021619. R.H. and A.J.L. are partly supported by the National Science Foundation under Award No. PHY-2114024.
\end{acknowledgments}

%----------------------------------------------------------------
% Supplementary Material
%----------------------------------------------------------------
\appendix
\onecolumngrid

\section{Calculation of axion emissivity}
\label{app:calculation_detail}

In this section, we implement the Fermi surface approximation and evaluate the axion emissivity from the process $l+f\rightarrow l'+f+a$, where a lepton $l=e,\mu$ is converted to another $l'=\mu,e$ with the spectator particle $f$ being one of $p, e, \mu$. This approximation was also used in the calculation of neutrino emissivities \cite{Friman:1979ecl, Yakovlev:2000jp} and axion emissivities for the bremsstrahlung process by nucleons \cite{Iwamoto:1984ir, Brinkmann:1988vi, Iwamoto:1992jp, Harris:2020qim}. The metric signature is $(-,+,+,+)$. 

The axion emissivity is calculated as
\begin{equation}
\label{eq:app-emissivity-integral}
\begin{split}
	\ve_a^{(lf)} & = 
     \frac{1}{S} 
    \int \! \frac{d^3p_1}{(2\pi)^3} \frac{1}{2E_1} 
    \frac{d^3p_2}{(2\pi)^3} \frac{1}{2E_2} 
    \frac{d^3p_1^\prime}{(2\pi)^3} \frac{1}{2E_1^\prime} 
    \frac{d^3p_2^\prime}{(2\pi)^3} \frac{1}{2E_2^\prime}  
    \frac{d^3p_3^\prime}{(2\pi)^3} \frac{1}{2E_3^\prime} 
    \sum\limits_\rm{spin} \, \bigl| \mathcal{M}^{(lf)} \bigr|^2 
    \\ & \quad 
    \times 
    (2\pi) \, \delta(E_1+E_2-E_1'-E_2'-E_3') \ 
    (2\pi)^3 \, \delta^{(3)}(\pvec_1+\pvec_2-\pvec_1'-\pvec_2'-\pvec_3') 
    \\ & \quad 
	\times E_3' \, f_1 \, f_2 \, (1-f_1') \, (1-f_2') ~,
\end{split}
\end{equation}
where $\mathcal{M}^{(lf)}$ is the Lorentz-invariant matrix element for the scattering $l + f \to l' + f + a$.  
The symmetry factor $S$ is needed to avoid double counting of identical particles if $l$ or $l' = f$. 
The energies $E_i$ are determined by the on-shell conditions: $E_i = \sqrt{|\pvec_i|^2 + m_i^2}$ for $i = 1, 2, 1', 2', 3'$.  
The thermal factors $f_1 \, f_2 \, (1-f_1') \, (1-f_2')$ restrict the fermion particle energies ($E_1$, $E_2$, $E_1'$, and $E_2'$) to be near their respective Fermi energies $E_{F,i}$ within a narrow range of order temperature $T \ll E_{F,i}$. 
This observation motivates the Fermi surface approximation, by which the emissivity is factorized into angular integrals with momenta restricted to the Fermi surface and energy integrals.  
To implement the Fermi surface approximation we introduce Dirac delta functions that fix the magnitude of the fermion 3-momenta to equal their respective Fermi momenta, and we promote the fermion energies to integration variables via the prescription: 
\begin{align}
	d^3p \rightarrow d^3p \int \frac{E}{p_F} \delta(p-p_F) dE ~.
\end{align}
This approximation allows the emissivity to be written as 
\begin{align}
	\ve_a^{(lf)} = \frac{1}{2^5(2\pi)^{11} p_{F,1} p_{F,2} p_{F,1'} p_{F,2'} S} J A ~,
\end{align}
which splits the calculation into two parts: an angular integral $A$ and an energy integral $J$, defined by 
\begin{align}
	\label{angular_integral}
	A & \equiv \int \! d^3p_1 d^3p_2 d^3p_1' d^3p_2' d^2\Omega_3' \delta(p_1-p_{F,1}) \delta(p_2-p_{F,2}) \delta(p_1'-p_{F,1'}) \delta(p_2' - p_{F,2'}) \delta^3(\b p_1 + \b p_2 - \b p_1' - \b p_2') \frac{\sum_\rm{spin} \abs{\cal M^{(lf)}}_\rm{Fermi}^2}{{E_3'}^{n}} ~,\\
	\label{energy_integral}
	J & \equiv \int \! dE_1 dE_2 dE_1' dE_2' dE_3' \delta(E_1+E_2-E_1'-E_2'-E_3') f_1 f_2 (1-f_1')(1-f_2') {E_3'}^{n+2} ~.
\end{align}
The matrix element $\abs{\cal M^{(lf)}}_\rm{Fermi}$ is evaluated with fermion 3-momenta and energies fixed to the respective Fermi momenta and Fermi energies.  
The exponent $n$ is chosen such that ${E_3'}^{-n}\sum_\rm{spin} \abs{\cal M^{(lf)}}_\rm{Fermi}^2$ is independent of $E_3'$. 
We have neglected the axion momentum in the momentum conservation delta function since $p_3'\sim T \ll p_{F,\mu}$. The mass dimension of $J$ and $A$ is $6+n$ and $3-n$, and that of $\abs{\cal M^{(lf)}}^2$ is $-2$. For the LFV channels considered in this work, we note that $p_{F,2}=p_{F,2'}$, $n=2$, and $S=1$ for $f$ being a proton and $S=2$ otherwise. 

\subsection{Energy integral}
The energy integral can be written as
\begin{align}
	J \approx \int_{-\infty}^\infty \! dx_1 \int_{-\infty}^\infty dx_2 \int_{-\infty}^\infty \! dx_1' \int_{-\infty}^\infty \! dx_2' \int_0^\infty \! dz \frac{T^{6+n} z^{2+n} \delta\( x_1 + x_2 + x_1' + x_2' - z\)}{(e^{x_1}+1) (e^{x_2}+1) (e^{x_1'}+1) (e^{x_2'}+1)} = \frac{T^{6+n}}{6} \int_0^\infty dz \frac{z^{3+n}(z^2+4\pi^2)}{e^z - 1} ~,
\end{align}
where 
$x_i \equiv (E_i-E_{F,i})/T$ , 
$x_i' \equiv (E_{F,i}'-E_i')/T$, 
and $z\equiv E_3'/T$.  
The approximation symbols arise from extending the limits of integration to infinity. 
The second equality is derived using the technique in \cite{baym2008landau}. 
For $n=2$, we obtain
\begin{align}
	J = \frac{164\pi^8}{945} T^8 ~.
\end{align}

\subsection{Angular integral}
For the angular integral, we first integrate $d^3p_2'$ with the momentum delta function and $dp_1, dp_2, dp_1'$ with the Fermi surface delta function. It is convenient to align all angles with respect to $\b p_1$, so $\int \! d^2\Omega_1$ simply gives $4\pi$. The angular integral $A$ becomes
\begin{align}
	\label{A1}
	\nonumber
	A &= 4\pi p_{F,1}^2 p_{F,2}^2 p_{F,1'}^2 \int_{-1}^1 dc_{12} \int_{-1}^1 dc_{11'} \int_{-1}^1 dc_{13'} \int_0^{2\pi} d\j_{12} \int_0^{2\pi} d\j_{11'} \int_0^{2\pi} d\j_{13'} ~ \delta(p_2'-p_{F,2'}) {E_3'}^{-n} \sum_\rm{spin} \abs{\cal M^{(lf)}}^2_\rm{Fermi} ~,\\
	&= 32\pi^3 p_{F,1}^2 p_{F,2}^2 p_{F,1'}^2 \int_{-1}^1 dc_{12} \int_{-1}^1 dc_{11'} \int_{-1}^1 dc_{13'} \int_0^\pi dv_\j ~ \delta(p_2'-p_{F,2'}) \la {E_3'}^{-n} \sum_\rm{spin} \abs{\cal M^{(lf)}}^2_\rm{Fermi} \ra _{\j_{13'}} ~,
\end{align}
where $c_{ij}$ denotes the cosine of the angle between $\b p_i$ and $\b p_j$, $u_\j\equiv \j_{11'} + \j_{12}$, $v_\j \equiv \j_{11'}-\j_{12}$, and $\la \cdots \ra_{\j_{13'}}$ stands for an average over $\j_{13'}$. To obtain the second equality, we have assumed that $\la {E_3'}^{-n} \sum_\rm{spin} \abs{\cal M}^2_\rm{Fermi} \ra _{\j_{13'}}$ and $\delta(p_2'-p_{F,2'})$ do not depend on $u_\j$, and may rely on $v_\j$ only through $\cos v_\j$.

To simplify the expression further, we note that $2$ and $2'$ represent identical particle species whereas $1$ and $1'$ represent different particle species, and either $p_{F,2}\geq p_{F,1},p_{F,1'}$ or $p_{F,2}< p_{F,1},p_{F,1'}$. The delta function then becomes
\begin{align}
	\label{A2}
	\delta(p_2'-p_{F,2'}) = \frac{ \delta( v_\j - v_{\j,0} ) }{ p_{F,1'} \sqrt{(1-c_{11'}^2) (1-c_{12}^2) (1-\cos^2 v_{\j,0}) } }  ~,
\end{align}
where
\begin{align}
	\label{A3}
	v_{\j,0} = \arccos\[ \frac{p_{F,1}^2 + p_{F,1'}^2 - 2 p_{F,1} p_{F,1'} c_{11'} + 2 p_{F,2} (p_{F,1} - p_{F,1'} c_{11'}) c_{12} }{2 p_{F,1'} p_{F,2} \sqrt{(1-c_{11'}^2) (1-c_{12}^2) } } \]~.
\end{align}
To have a real-valued $v_{\j,0}$ within the range from $0$ to $\pi$, we must require $\cos^2 v_{\j,0} < 1$. This restricts the range of $dc_{11'}$ and $dc_{12}$ integrals to be within
\begin{align}
	\label{A4}
	c_{11'}^- < c_{11'} < c_{11'}^+ \sep
	c_{12}^- < c_{12} < c_{12}^+ ~,
\end{align}
where
\begin{align}
	\label{A5}
	c_{11'}^\pm &= \frac{(p_{F,1}+p_{F,2} c_{12}) (p_{F,1}^2 + p_{F,1'}^2 + 2 p_{F,1} p_{F,2} c_{12}) }{2 p_{F,1'} (p_{F,1}^2 + p_{F,2}^2 + 2p_{F,1} p_{F,2} c_{12})}  \pm \frac{ p_{F,2} \sqrt{(c_{12}^2-1) [(p_{F,1}^2 - p_{F,1'}^2 + 2 p_{F,1} p_{F,2} c_{12})^2 - (2p_{F,2} p_{F,1'} )^2] }}{2 p_{F,1'} (p_{F,1}^2 + p_{F,2}^2 + 2p_{F,1} p_{F,2} c_{12})} ~,
\end{align}
and
\begin{align}
	\label{A6}
	c_{12}^+ = \min\[1, \frac{p_{F,1'}^2 - p_{F,1}^2 + 2 p_{F,2} p_{F,1'}}{2 p_{F,1} p_{F,2}} \] \sep
	c_{12}^- = \max\[-1, \frac{p_{F,1'}^2 - p_{F,1}^2 - 2 p_{F,2} p_{F,1'}}{2 p_{F,1} p_{F,2}} \] ~.
\end{align}
Combining equations \eqref{A1}-\eqref{A6}, we find
\begin{align}
	\label{lfv_angular_integral8}
	A = 32\pi^3 p_{F,1}^2 p_{F,2}^2 p_{F,1'} \int_{c_{12}^-}^{c_{12}^+} dc_{12} \int_{c_{11'}^-}^{c_{11'}^+} dc_{11'} \int_{-1}^1 dc_{13'} \frac{ \la {E_3'}^{-n} \sum_\rm{spin} \abs{\cal M^{(lf)}}^2_\rm{Fermi} \ra _{\j_{13'}, v_\j = v_{\j,0}} }{ \sqrt{(1-c_{11'}^2) (1-c_{12}^2) (1-\cos^2 v_{\j,0}) } } ~.
\end{align}
We need to calculate the matrix element at the Fermi surface to evaluate this integral.

\subsection{Matrix element}
\begin{figure}
	\centering
	\includegraphics[width=0.7\linewidth]{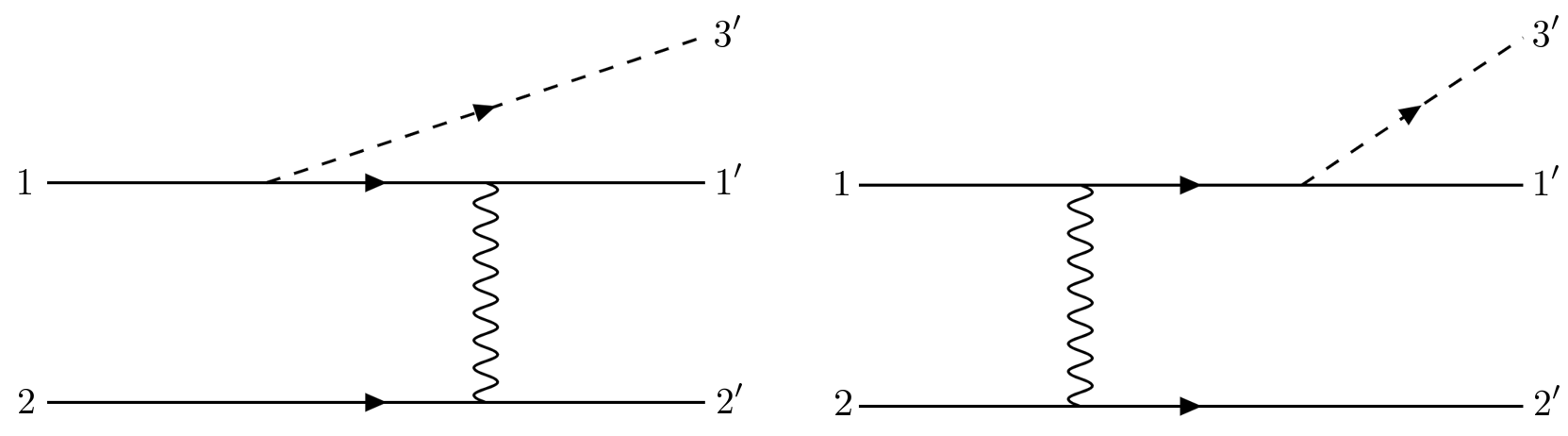}
	\caption{Feynman diagrams for the LFV process $l + f \rightarrow l' + f + a$. If $f$ is a lepton, there occur two more graphs which can be obtained by exchanging $(1\leftrightarrow 2)$ for $f$ being identical to $l$ or $(1'\leftrightarrow 2')$ for $f$ being identical to $l'$.}
	\label{fig:lfvfeynmandiagram}
\end{figure}

Now we evaluate the matrix element. It is convenient to use the LFV coupling
\begin{align}
	\cal L_\rm{LFV} = -i g_{ae\mu} a ( \bar\Psi_e \gamma_5 \Psi_\mu + \bar\Psi_\mu \gamma_5 \Psi_e ) ~,
\end{align}
which is equivalent to the use of the pseudovector (derivative) form written in the main text if each fermion line is attached to at most one axion line \cite{Raffelt:1987yt}. Given the two Feynman diagrams in figure \ref{fig:lfvfeynmandiagram}, the matrix elements are
\begin{align}
	i\cal M^{(1)} = \pm e^2 g_{ae\mu} \[ \bar u_1' \gamma^\mu \frac{-\sl r + m_1'}{r^2 + {m_1'}^2} \gamma_5 u_1 \] \frac{-g_{\mu\nu}}{k^2} [\bar u_2' \gamma^\nu u_2] ~,\\
	i\cal M^{(2)} = \pm e^2 g_{ae\mu} \[ \bar u_1' \gamma_5 \frac{-\sl s + m_1}{s^2 + m_1^2} \gamma^\mu  u_1 \] \frac{-g_{\mu\nu}}{k^2} [\bar u_2' \gamma^\nu u_2] ~,
\end{align}
where $k\equiv p_2-p_2'$, $r\equiv p_1-p_3'$, $s\equiv p_1'+p_3'$ and $\pm$ refers to the sign of the spectator particle's electric charge. In NSs we have $|m_1^2 - {m_1'}^2| \approx m_\mu^2 \gg E_F E_3'$, thus $r^2 + {m_1'}^2 \approx -m_1^2 + {m_1'}^2$ and $s^2 + m_1^2 \approx -{m_1'}^2 + m_1^2$. The matrix element for exchange diagrams can be obtained by $(1\leftrightarrow 2)$ or $(1'\leftrightarrow 2')$, with an additional factor of $-1$ included. 

The spin-summed squared matrix element is
\begin{align}
    \label{Mlp}
	\sum_\rm{spin} \abs{\cal M^{(lp)}}^2 &= -\frac{128 g_{ae\mu}^2 e^4}{(p_2-p_2')^4} \frac{(p_1\cdot p_1' + m_1 m_1') (p_2\cdot p_3') (p_2'\cdot p_3')}{(m_1^2 - {m_1'}^2)^2} ~, \\
    \label{Mll}
    \sum_\rm{spin} \abs{\cal M^{(ll)}}^2 &= \sum_\rm{spin} \abs{\cal M^{(lp)}}^2 + (1\leftrightarrow 2) + \cal T^{(ll)} ~, \\
    \label{Mll'}
    \sum_\rm{spin} \abs{\cal M^{(ll')}}^2 &= \sum_\rm{spin} \abs{\cal M^{(lp)}}^2 + (1'\leftrightarrow 2') + \cal T^{(ll')} ~,
\end{align}
where $l=e,\mu$ and $l'=\mu,e$. The second term in \eqref{Mll} and \eqref{Mll'} is the contribution solely from the exchange diagrams given by the first term but with $(1\leftrightarrow 2)$. The third term in \eqref{Mll} is the interference between prototype and exchange diagrams given by
\begin{align}
	\nonumber
	\cal T^{(ll)} &= \frac{64 g_{ae\mu}^2 e^4}{(p_1-p_2')^2 (p_2-p_2')^2} \frac{p_2' \cdot p_3'}{(m_1^2 - {m_1'}^2)^2} \\
	&\phantom{=}\times [(p_2\cdot p_1' + m_1 m_1') (p_1\cdot p_3') + (p_1\cdot p_1' + m_1 m_1') (p_2\cdot p_3') - (p_1\cdot p_2 + m_1^2) (p_1'\cdot p_3')] ~,
    \label{Tll}
\end{align}
and $T^{(ll')}$ in \eqref{Mll'} by $\cal T^{(ll)}$ but with $(1\leftrightarrow 1')$ and $(2\leftrightarrow 2')$. Here we evaluate the traces of products of gamma matrices and spinors with the help of the Mathematica package FeynCalc \cite{Shtabovenko:2020gxv}.

At the Fermi surface, the spin-summed squared matrix element becomes
\begin{align}
	\sum_\rm{spin} \abs{\cal M^{(lf)}}^2_{\rm{Fermi}} = \frac{32 e^4 g_{ae\mu}^2 {E_3'}^2}{E_{F,1}^2 E_{F,2}^2 \beta_2^4 (\beta_1^2 - {\beta_1'}^2)^2} G^{(lf)} ~,
\end{align}
where $f=p,e,\mu$. The $G^{(lf)}$ factor is found to be
\begin{align}
    \label{Glp}
	G^{(lp)} &= \frac{(1-\beta_{F,2} c_{23'}) (1-\beta_{F,2} c_{2'3'}) (1- \beta_{F,1} \beta_{F,1'} c_{11'})}{(1-c_{22'})^2} ~,\\
    \label{Gll}
    G^{(ll)} &= G^{(lp)} + (1\leftrightarrow 2) + H^{(ll)} ~,\\
    \label{Gll'}
    G^{(ll')} &= G^{(lp)} + (1'\leftrightarrow 2') + H^{(ll')} ~,
\end{align}
where we have assumed that electrons are ultra relativistic so $\beta_{F,e}=1$. The second term in \eqref{Gll} and \eqref{Gll'} is the contribution solely from the exchange diagrams given by the first term but with $(1\leftrightarrow 2)$. The third term in \eqref{Gll} is the interference between prototype and exchange diagrams given by
\begin{align}
	\nonumber
	H^{(ll)} = \frac{(1-\beta_{F,1} c_{2'3'}) }{2 (1-c_{12'}) (1-c_{22'})} &[  \beta_{F,1}\big( c_{13'}+c_{23'} + \beta_{F,1}(1-c_{12}) + \beta_{F,1'}(c_{11'} + c_{21'}) \\
	& + \beta_{F,1}\beta_{F,1'} ( c_{12} c_{1'3'} - c_{11'} c_{23'} - c_{13'} c_{21'} - c_{1'3'} ) \big) - 2 ] ~,
\end{align}
and $H^{(ll')}$ in \eqref{Gll'} by $H^{(ll)}$ but with $(1\leftrightarrow 1')$ and $(2\leftrightarrow 2')$.

\subsection{Axion emissivity}
\begin{figure}
	\centering
    \begin{minipage}{0.49\linewidth}
		\includegraphics[width=\linewidth]{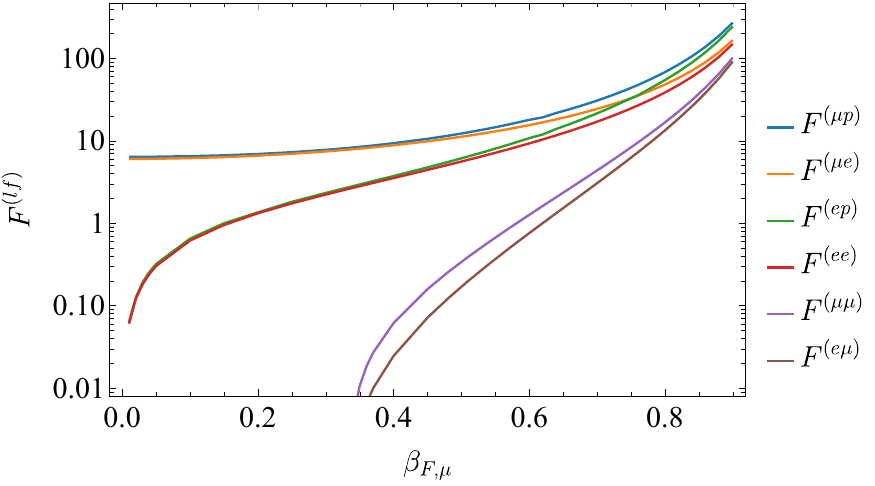}
	\end{minipage}\quad
	\begin{minipage}{0.49\linewidth}
		\includegraphics[width=\linewidth]{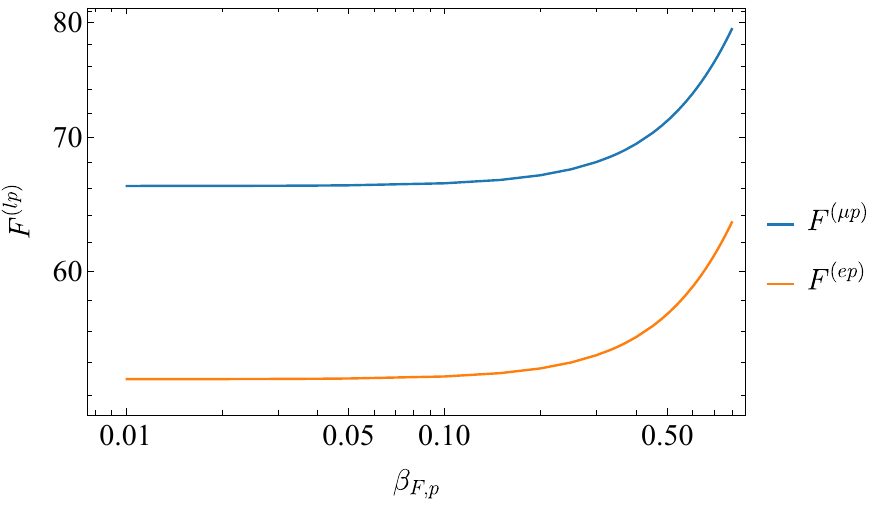}
	\end{minipage}
	\caption{The factor $F^{(lf)}$ as a function of the Fermi velocity of muons (left) and protons (right). Here we have set $\beta_{F,p}=0.3$ and $\beta_{F,\mu}=0.8$ for the left and right panels respectively for the $f=p$ processes.}
	\label{fig:lfvf}
\end{figure}

In summary, the axion emissivity is given by
\begin{align}
    \label{axion_emissivity}
	\ve_a^{(lf)} &= \frac{328\pi^2 \alpha^2 g_{ae\mu}^2 }{945 m_\mu^4} \frac{\beta_{F,1} E_{F,1}^3 }{\beta_{F,2}^2 p_{F,2}^2} F^{(lf)} T^8 ~,\\
	F^{(lf)} &\equiv \frac{1}{8S} \int_{c_{12}^-}^{c_{12}^+} dc_{12} \int_{c_{11'}^-}^{c_{11'}^+} dc_{11'} \int_{-1}^1 dc_{13'} \frac{ \la G^{(lf)}\ra_{\j_{13'}, v_\j = v_{\j,0}} }{ \sqrt{(1-c_{11'}^2) (1-c_{12}^2) (1-\cos^2 v_{\j,0}) } } ~.
\end{align}
The $dc_{13'}$ integral can be evaluated analytically. 
We calculate the other integrals using numerical techniques and present the result for $F^{(lf)}$ in figure \ref{fig:lfvf}. 
In the left panel we vary the muon Fermi velocity $\beta_{F,\mu} = p_{F,\mu} / E_{F,\mu}$.  
From the right panel we see that $F^{(lp)}$ is not sensitive to $\beta_{F,p}$ if protons are nonrelativistic, i.e., $\beta_{F,p} \lesssim 0.5$, which is expected in NSs. 
Therefore, we use the values of $F^{(lf)}$ shown in the left panel to calculate the emissivity shown in the main text.

\subsection{Different temperature dependence from LFV and LFP interactions}
In the main text we contrast the temperature dependence of the axion emissivity for LFV and LFP interactions. 
The LFP interaction leads to axion emission via channels such as $l + f \to l + f + a$ with an emissivity that scales as $\ve_a \propto T^6$ (similar for $nn \to nna$ \cite{Raffelt:1990yz}).  
By considering the LFV interaction here, we find that channels such as $l + f \to l' + f + a$ lead to an emissivity $\ve_a \propto T^8$ instead.  
This different scaling may be understood by inspecting the form of the matrix element.  
Consider the Feynman diagram in the left panel of figure \ref{fig:lfvfeynmandiagram}. 
The fermion propagator and the axion vertex contribute factors of 
\begin{align}
    \frac{E_3'}{(p_1-p_3')^2+m_1^{\prime\,2
}} = \frac{E_3'}{m_1^{\prime\,2}-m_1^2 + 2E_3'(E_1 - 2 |\pvec_1| c_{13'})} ~,
\end{align}
in the $(-,+,+,+)$ metric signature and neglecting the axion mass $E_3' = |\pvec_3'|$.
The axion energy $E_3'$ in the numerator arises from the derivative nature of the axion interaction. 
The temperature dependence enters via the typical axion energy, $E_3' \sim T$.  
For LFP channels such as $\mu p \rightarrow \mu p a$, we have $m_1^\prime = m_1$, the $E_3^\prime \sim T$ factor in the numerator is canceled by the factor in the denominator, and consequently the squared matrix element is insensitive to the temperature. 
On the other hand, for the LFV channels, the $m_1^{\prime\,2}-m_1^2$ term dominates in the denominator.  
Consequently, the LFV axion emissivity is suppressed relative to the LFP calculation by a factor of order $T^2 E_{F,e}^2 / (m_\mu^2 - m_e^2)^2 \sim T^2/m_\mu^2 \sim 7 \times 10^{-7} T_9^2$.

\section{The \texorpdfstring{$npe\mu$}{npeu} matter}
\label{app:npeu_matter}
At typical NS densities $\sim 10^{15} \, \rm{g~cm^{-3}}$, the equilibrium composition involves neutrons, protons, electrons, muons and other exotic matter states such as hyperons. Neglecting the exotic matter, equations of state for a NS are relatively easy to calculate \cite{cohen1970neutron}. Thermal equilibrium and conservation of the baryon number and electric charge impose \cite{Haensel:2007yy}
\begin{align}
	E_{F,\mu} = E_{F,e} \sep
	E_{F,n} = E_{F,p} + E_{F,e} \sep
	n_p = n_e + n_\mu ~,
\end{align}
where we have approximated the chemical potential with the Fermi energy. We also have the Fermi energy $E_{F,i}^2 = m_i^2 + p_{F,i}^2$, the number density $n_i = p_{F,i}^3 / 3 \pi^2$, and the mass density $\rho = \sum_i m_i n_i$. If one of $\rho, n_n, n_p, n_e, n_\mu$ is fixed, the other quantities can be fully determined. For this work, we have taken $0.8 \, m_N \approx 750 \MeV$ for the mass of nucleons to account for their nuclear interactions. At $\rho = 6\rho_0 \approx 1.5 \times 10^{15} \, \mathrm{g~cm^{-3}}$, we find
\begin{align}
\label{fermi_momenta_npemu}
    p_{F,n} \simeq 624 \MeV \sep
    p_{F,p} \simeq 226 \MeV \sep
	p_{F,e} \simeq 193 \MeV \sep
	p_{F,\mu} \simeq 162 \MeV ~,
\end{align}
corresponding to $\beta_{F,p}\simeq 0.29$ and $\beta_{F,\mu}\simeq 0.84$.

\section{Numerical integration}
\label{app:numerical-integration}

\subsection{Numerical integrator}
In this section we discuss the numerical method used to evaluate \eqref{eq:app-emissivity-integral}. 
To prepare the integrand for numerical integration we simplify it by using the Dirac deltas to perform $4$ integrals analytically. 
We use the momentum conserving Dirac delta to carry out the $d^3 p_2'$ integrals which enforces $\bm{p}_2' = \bm{p}_1 + \bm{p}_2 - \bm{p}_1' - \bm{p}_3'$.
Next, we rewrite the momentum integrals in spherical coordinates by making the replacements $d^3 p \rightarrow |\bm{p}|^2 \, d |\bm{p}| \, d \cos \theta \, d \phi$ where $\theta$ and $\phi$ give the polar and azimuthal angles of $\bm{p}$ in the rest frame of the NS. 
The coordinate system is oriented so that the $z$-axis points in the same direction as $\bm{p}_3'$ so that the $d \cos \theta_3' \, d \phi_3'$ integral yields a trivial factor of $4 \pi$.
We then change variables from momentum magnitudes $|\bm{p}|$ to energies by using the relation $E^2 = |\bm{p}|^2 + m^2$ to write $E \, d E = |\bm{p}| \, d |\bm{p}|$.
Finally, the energy Dirac delta is used to fix $|\bm{p}_3'|$ so that, assuming the axion is massless ($m_3' = 0$),
\begin{equation}
\begin{aligned}
    &E_1 + E_2 - E_1' - E_2' - E_3' = E_1 + E_2 - E_1' - \sqrt{|\bm{p}_1 + \bm{p}_2 - \bm{p}_1' - \bm{p}_3'|^2 + m_2'^2} - |\bm{p}_3'| \\
    &=
    \label{eq:app-root}
    E_1 + E_2 - E_1' - \sqrt{|\bm{P}|^2 - 2 \, P_z \, |\bm{p}_3'| + |\bm{p}_3'|^2   + m_2'^2} - |\bm{p}_3'| = 0  
\end{aligned}
\end{equation}
where $\bm{P} \equiv \bm{p}_1 + \bm{p}_2 - \bm{p}_1' - \bm{p}_3'$.
This adds a factor of $|1 + (|\bm{p}_3'| - P_z) / E_2|^{-1}$ to the integrand since $\delta[f(x)] = \delta(x - x_*) / |f'(x_*)|$ where $x_*$ is the root of $f(x)$. 
In practice, \eqref{eq:app-root} is enforced by using Newton-Raphson iteration to find the value of $|\bm{p}_3'|$ which is a root of this equation when the integration variables $E_1,\,E_2,\,E_1',\,\cos \theta_1,\,\cos \theta_2,\,\cos\theta_1',\,\phi_1,\,\phi_2,$ and $\phi_1'$ are fixed.
All together, this rewrites the integral \eqref{eq:app-emissivity-integral} as
\begin{equation}
\label{eq:app-emissivity-integral-numerical}
\begin{split}
	\ve_a^{(lf)} = 
     \frac{4 \pi}{2^5 (2\pi)^{11}} \frac{1}{S} 
    \int \! 
    & d E_1 \, d \cos \theta_1 \, d \phi_1 \,
    d E_2 \, d \cos \theta_2 \, d \phi_2 \,
    d E_1' \, d \cos \theta_1' \, d \phi_1' \\
    &\quad \times \frac{
        |\bm{p}_1|
        |\bm{p}_2|
        |\bm{p}_1'|
        |\bm{p}_3'|
    }
    {
        E_2' \, 
        \bigl|
            1 + (E_3' - P_z) / E_2'
        \bigr|
    } 
    \sum\limits_\rm{spin} \, \bigl| \mathcal{M}^{(lf)} \bigr|^2 \, E_3' \, f_1 \, f_2 \, (1-f_1') \, (1-f_2') ~,
\end{split}
\end{equation}
where the matrix element is given by (\ref{Mlp} - \ref{Mll'}).
We evaluate the integral in this form using the Vegas package in Python which performs Monte Carlo integration using two adaptive strategies: importance sampling, and stratified sampling,  to improve convergence~\cite{Lepage:2020tgj}.
We choose to use this Monte Carlo integrator because of its flexibility and ease of use. 
The integral is evaluated by passing the integrand as an explicit function of the $9$ integration variables $(E_1, E_2, E_1', \cos \theta_1, \cos \theta_2, \cos \theta_1', \phi_1, \phi_2, \phi_1')$ to an instance of the \texttt{vegas.Integrator} class. 
We split the calculation of the integral into two steps. 
First, we adapt the \texttt{vegas.Integrator} object to the integrand by calling it with the parameters \texttt{nitn} $=10$, \texttt{neval} $= 5 \times 10^7$, and \texttt{alpha} $=0.1$.
These parameters control the number of iterations used to adapt the integrator; the number of points on the integration domain where the integrand is evaluated; and the sensitivity of the adaptation algorithms, respectively.
We then discard the results obtained from the first run but keep the adapted integrator and call it again with the same parameter choices except with \texttt{alpha = 0} so that there is no further adaptation.
The value of the integral and the errors we report below are taken as the \texttt{mean} and \texttt{sdev} attributes of the second run \texttt{vegas.Integrator} object. 
The mean is a weighted average of the results of each of the \texttt{nitn = 10} iterations of the Vegas algorithm, where the weights are the inverse variance in each iteration. 
The uncertainty, \texttt{sdev} is the square root of the variance of the weighted average assuming the sample average in each iteration is approximately normally distributed -- this is a good approximation if \texttt{neval} is sufficiently large. 

\begin{figure}
    \centering
    \includegraphics[width=0.85\linewidth]{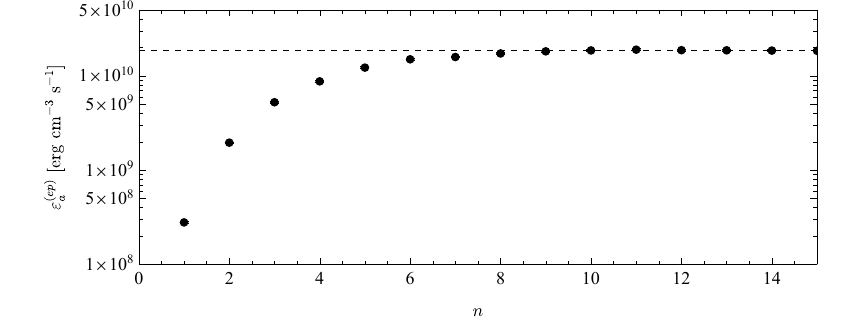}
    \caption{
        Axion emissivity for the $ep \rightarrow \mu p a$ channel vs energy integration domain $E_i \in [E_{F,i} - n T,\ E_{F,i} + n T]$ parameterized by $n$.
        The gray dashed line is the constant $1.85 \times 10^{10} \, \mathrm{erg} \, \mathrm{cm}^{-3} \, \mathrm{s}^{-1}$, which is the value to which the integral converges.
        For these calculations we have fixed $\beta_{F,\mu} = 0.836788$, $g_{ae\mu} = 10^{-11}$, and $T = 10^9\,\mathrm{K}$. 
        As $n$ increases, the value of the emissivity integral converges to a constant value of $\approx 1.8 \times 10^{10} \, \mathrm{erg} \, \mathrm{cm}^{-3} \, \mathrm{s}^{-1}$ . 
    }
	\label{fig:convergence-check}
\end{figure}

In principle the energy integrals over $E_1$, $E_2$, $E_1'$ should be over the domain $E_i \in [m_i, \infty)$ but in practice we can only integrate over a finite window.
The thermal factors in \eqref{eq:app-emissivity-integral-numerical} provide support only in a window around the Fermi level $E_{F} = \sqrt{p_F^2 + m^2}$ whose width is of order $\sim T$.
This motivates integrating $E_1$, $E_2$, $E_1'$ over the finite window $E_i \in [\mathrm{max}(m_i, E_{F,i} - n T), \, E_{F,i} + n T]$ with a value of $n$ sufficiently large that the integral is insensitive to its exact value.
We find $n = 10$ to be large enough that the integral is independent of $n$, but small enough that Monte Carlo convergence is not too slow. 
The $n$-independence is demonstrated for the process $ep \rightarrow \mu p a$ for $\beta_{F,\mu} = 0.84$, $T = 10^9~\mathrm{K}$ and $m_a = 0$ in figure \ref{fig:convergence-check}.
Note how  as $n$ increases, the emissivity approaches a constant value of approximately $1.8 \times 10^{10}~\mathrm{erg}~\mathrm{cm}^{-3}~\mathrm{s}^{-1}$, which corresponds to the blue data point at $\beta_{F,\mu} \approx 0.84$ in figure \ref{fig:numerical-emissivity}.

\subsection{Numerical validation of Fermi surface approximation}
\begin{figure}
    \centering
    \includegraphics[width=0.55\linewidth]{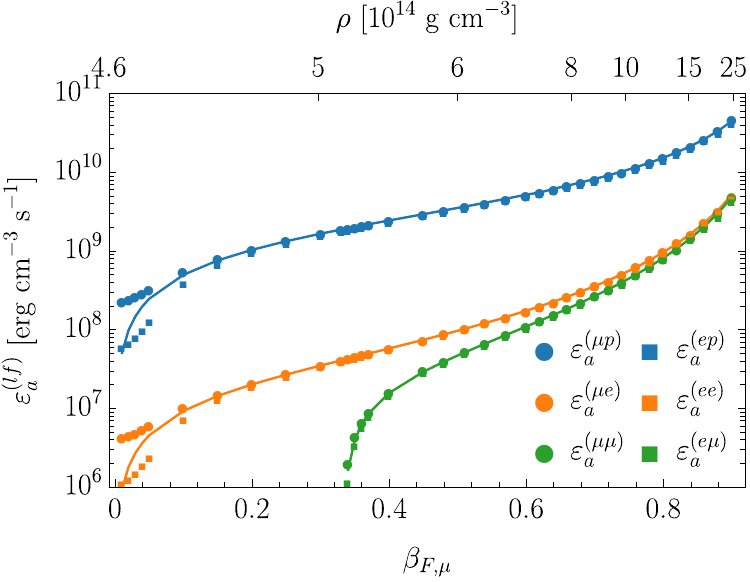}
    \caption{
        Axion emissivity computed using the Monte Carlo integration method (dots and squares) vs. Fermi surface approximation (lines). 
        The results agree well for $\beta_{F,\mu} \gtrsim 0.1$ and the agreement is good within about 10\% at $\beta_{F,\mu} \approx 0.8$. 
        At small $\beta_{F,\mu}\lesssim 0.1$, the Fermi surface approximation underestimates the emissivity for the $\mu p$, $\mu e$ channels and overestimates it for the $ep$ and $ee$ channels. To make this plot, we choose $g_{ae\mu} = 10^{-11}$ and $T = 10^9~\Kel$ to be consistent with the parameters in figure 2 of the main text.
    }
	\label{fig:numerical-emissivity}
\end{figure}
The results of our numerical evaluations of \eqref{eq:app-emissivity-integral-numerical} for the various axion emission channels are shown in figure \ref{fig:numerical-emissivity}.
The numerical results (dots and squares) agree very well with the analytical results (lines) for a wide range of $\beta_{F,\mu}$. 
For small $\beta_{F,\mu}$ the numerical results tend to diverge from the analytical results, which is expected because in this regime the number density of muons is small, which means that the degenerate matter approximation breaks down.
In addition, we observe that for $\beta_{F,\mu}\gtrsim 0.1$ the emissivities are paired by channel such that $\ve_a^{(lf)}\approx \ve_a^{(l'f)}$. This is a consequence of the strong particle degeneracy and the beta equilibrium condition $E_{F,e}\approx E_{F,\mu}$.
We have verified this numerically and analytically by imposing the relation $E_{F,e} = E_{F,\mu} + \Delta$ and observing that the difference between emissivities for the channels $ef \rightarrow \mu f a$ and $\mu f \rightarrow e f a$  grows with $\Delta$ but is only significant if $\Delta \gtrsim T$. 
For $\Delta > 0$ the electron's Fermi energy is larger than the muon's which allows for electrons with energies below the Fermi level to also convert into muons, enhancing the emissivity of this channel.
Conversely, for muon to electron conversion, the emissivity is exponentially suppressed since the muons' energies are below the electrons' energies.

\subsection{Effect of temperature on axion emissivity}
In addition to verifying that numerically evaluating the axion emissivity at $T = 10^9~\mathrm{K}$ agrees with the analytical approximation, we also numerically computed the axion emissivity as a function of temperature while fixing $\beta_{F,\mu}=0.836788$ and $m_a = 0$.
We are motivated to do this for two reasons. The first is to confirm the $T^8$ scaling of the emissivity at low temperatures, i.e. equation \eqref{axion_emissivity}. The second is to calculate the emissivity for larger temperatures such as $T \sim 50~\mathrm{MeV}$, the scale of supernovae; allowing us to comment on constraints imposed on axion LFV interactions by supernovae observations.

In degenerate NS matter, there is a screening of electromagnetic fields due to the presence of a degenerate plasma with charged constituents. To estimate this effect, we replace the photon propagator $k^{-2}$ in the matrix element by $(k^2+ k_\rm{TF}^2)^{-1}$ \cite{Raffelt:1996wa}, where $k_\rm{TF}^2 = \sum_{i} 4\alpha p_{F,i} E_{F,i}/\pi$ is the Thomas-Fermi screening scale which receives contributions from electrons, muons and protons. Noting that $k^2 \sim (p_{F,e}-p_{F,\mu})^2 \sim E_{F,e}^2 (1-\beta_{F,\mu})^2$ at low temperatures, the screening effect is insignificant if $\beta_{F,\mu} \lesssim 1-k_\rm{TF}/E_{F,e}$, which becomes $\beta_{F,\mu} \lesssim 0.75$ at the core condition given by \eqref{fermi_momenta_npemu}. Therefore, for mildly relativistic muons with $\beta_{F,\mu}\sim 0.8$, the emissivity of LFV axions without including the screening effect is subject to $\cal O(1)$ corrections. On the other hand, incorporating the screening effect in axion emissivities is important at high temperatures since $k_\rm{TF}^2$ dominates over $k^2$, especially near the pole $k^2=0$.

\begin{figure}
    \centering
    \includegraphics[width=0.5\linewidth]{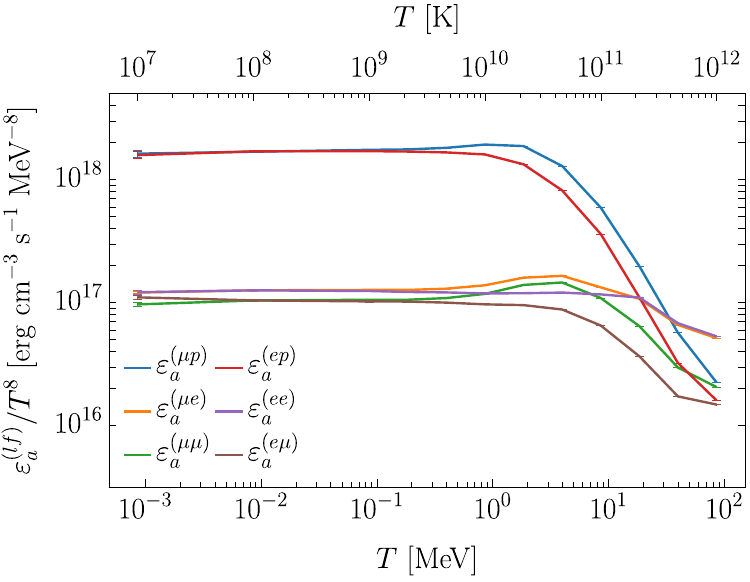}
    \caption{
        Numerically evaluated axion emissivity vs. temperature, calculated using \eqref{eq:app-emissivity-integral-numerical} with matrix elements given by (\ref{Mlp} - \ref{Mll'}). To generate these data we fixed $\beta_{F, \mu} = 0.836788$, $m_a = 0$, and $g_{ae\mu} = 10^{-11}$.
        The data presented here were computed with \texttt{neval} $= 5 \times 10^7$.
        The error bars are typically between $100$ to $10,\!000$ times smaller the mean values.
    }
	\label{fig:numerical-emissivity-vs-T}
\end{figure}

The temperature dependence of the axion emissivity is presented in figure \ref{fig:numerical-emissivity-vs-T} for $10^{-3}~\mathrm{MeV} \leq T \leq 100~\mathrm{MeV}$. 
Since we expect the emissivity to scale as $\varepsilon^{(lf)} \propto T^8$ for low temperatures we normalize the emissivity by $T^8$ so that a $T^8$ scaling would be a constant line in this figure.
The figure displays several interesting features.  
(1)  
At temperatures below $T \sim 10~\mathrm{MeV}$, the emissivity is seen to scale like $\varepsilon^{(lf)} \propto T^8$ (up to $\cal O(1)$ factors), which confirms the prediction from the Fermi surface approximation.
(2)  
The emissivity tends to decrease relative to $T^8$ for all six channels at temperatures $T \gtrsim 10 \MeV$.  
(3)  
For lower temperatures, the emissivities are paired by channel such that $\varepsilon_a^{(lf)} \approx \varepsilon_a^{(l'f)}$; however, at higher temperatures these relations do not hold. This is expected since the Fermi surface approximation, one of the assumptions needed to show that $\varepsilon_a^{(lf)}$ and $\varepsilon_a^{(l'f)}$ coincide, breaks down in this regime.
The significance of $T = 10 \MeV$ can be understood as follows: at low temperatures the thermal factors lead to a strong suppression of the integrand away from the Fermi surface.  As we lift the temperature the accessible phase space broadens and the pole becomes significant.

\subsection{Effect of axion mass on emissivity}
\begin{figure}
    \centering
    \includegraphics[width=0.5\linewidth]{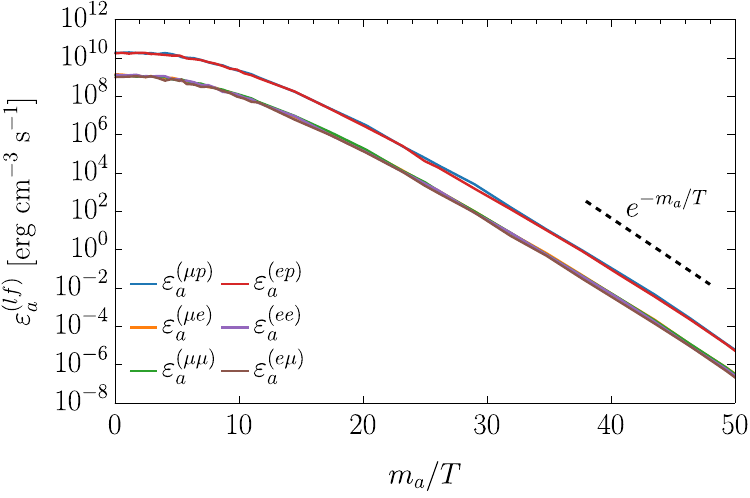}
    \caption{
        Numerically evaluated axion emissivity vs. axion mass, calculated using \eqref{eq:app-emissivity-integral-numerical} with matrix elements given by (\ref{Mlp} - \ref{Mll'}). 
        To generate these data we fixed $\beta_{F, \mu} = 0.836788$, $T = 10^9~\mathrm{K}$, and $g_{ae\mu} = 10^{-11}$.
        For large masses the emissivity falls off with an exponential tail (compare with black dashed line).
        The data presented here were computed with \texttt{neval} $= 10^6$.
    }
	\label{fig:numerical-emissivity-vs-ma}
\end{figure}
In previous results we assumed axions were massless. 
Here, we use our numerical integration method to explore the effect of raising the axion mass on the emissivity. 
To do this we must modify \eqref{eq:app-root} to accommodate a massive axion by replacing $E_3' = |\bm{p}_3'|$ with $E_3' = \sqrt{|\bm{p}_3'|^2 + m_a^2}$ so that energy conservation imposes the following constraint on $|\bm{p}_3'|$,
\begin{equation}
    \label{eq:app-root-massive-axion}
    E_1 + E_2 - E_1' - \sqrt{|\bm{P}|^2 - 2 \, P_z \, |\bm{p}_3'| + |\bm{p}_3'|^2   + m_2'^2} - \sqrt{|\bm{p}_3'|^2 + m_a^2} = 0 ~.
\end{equation}
In principle, we must also account for the axion's mass in the matrix element since \eqref{Mlp}--\eqref{Mll'} were derived assuming $m_a = 0$. 
However, we argue that the most important contribution of the mass to the emissivity is an exponential suppression arising from the thermal factors and therefore report results obtained using the `massless' matrix element of (\ref{Mlp} - \ref{Mll'}).
We set the temperature $T$ to a fiducial value of $10^9~\mathrm{K}$ and fix $\beta_{F,\mu} = 0.836788$ and calculate the emissivity for a range of masses satisfying $0 \leq m_a / T \leq 50$.
The emissivities calculated are presented in figure \ref{fig:numerical-emissivity-vs-ma}.
We find that the emissivity is approximately constant for $m_a / T\leq 10$, after which point the emissivity is exponentially suppressed. 

\twocolumngrid
\bibliographystyle{apsrev4-2}
\bibliography{ref}
\end{document}